\documentstyle[aps,prl,epsfig]{revtex}  
\begin{document} 
\draft 
\title{Nonlocal electrodynamics of two-dimensional  
 wire mesh photonic crystals}
\author{A. L. Pokrovsky, A. L. Efros}  
\address{University of Utah, Salt Lake City UT 84112 USA} 
\maketitle 

\begin{abstract} 
We calculate analytically the spectra of plasma waves and electromagnetic waves
(EMW) in metallic
photonic  crystal consisting of the parallel thin infinite  metallic cylinders
embedded in the dielectric media.  The axes of metallic cylinders form a
regular square lattice in a plane perpendicular to them. The metal inside the
cylinders is assumed to be in the high frequency regime $\omega \tau \gg 1$,
where $\tau$ is the relaxation time. The proposed analytical theory  is
based upon small parameters $f\ll 1$, where $f$ is the volume 
fraction of the metal, and
$kR\ll 1$, where $k$ is the wave vector and $R$ is the radius of the cylinder. 
It is
shown that there are five different branches of the  EMW that cover all 
frequency range under consideration except
one very small omnidirectional 
gap in the vicinity of the frequency of the surface plasmon.
However, at some directions of propagation and polarizations
the gap may be much larger.
The reflection and refraction of the EMW is also considered.
The general theory of refraction is proposed which is complicated 
by the spatial dispersion of the dielectric constant,
and one particular geometry of the incident EMW is considered.
\end{abstract}

\section{Introduction}

There has been a growing interest in recent years in the theoretical
and experimental study of the photonic crystals, 1-, 2- or 3-dimensional   
periodic structures that exhibit gaps for the propagation of  electromagnetic  
waves (EMW) in different frequency ranges. 
Those gaps are similar  to the Bragg's reflection of the X-rays\cite{book}. 
The important difference between the two effects 
is that in the X-ray theory one  
can use high frequency expansion of the dielectric constant to consider the  
interaction of the EMW with the media, while in the optical and microwave
range this is impossible. The physical origin of the difficulties  which may  
appear at low frequencies is the existence of other than EMW   
degrees of freedom. 
In  dielectric photonic crystals the polaritons might be important\cite{soc1}  
while in the metallic photonic crystals (MPC) these degrees of freedom are  
provided by free electrons. 
There are different ways of classification of the MPC. 
The first criterion is dimensionality, they may be 1-dimensional\cite{scarola},
2-dimensional\cite{mar}, and 3-dimensional\cite{yab1,soc2,joan,pen1,zak}.
The second criterion is the connectivity.
The MPC may consist of the separated metallic islands\cite{yab2}
or it may include a connected metallic mesh\cite{yab1,pen1}.
Both cases are considered theoretically in the review\cite{shalaev}.
The third criterion reflects the frequency dispersion of the metallic
conductivity.
This dispersion is negledgible if
$\omega \tau \ll 1$, where $\omega$ is the frequency of EMW and $\tau$
is a relaxation time of electrons in the metal.
In the opposite case, $\omega \tau \gg 1$, the dispersion of conductivity
is very important, 
and the dielectric constant of the metal can be written in the form
\begin{equation} 
\label{ep}
	\epsilon_p(\omega) = 1 -\frac{\omega_p^2}{\omega^2}, 
\end{equation}
where $\omega_p^2 = 4 \pi n e^2/m$ is the plasma frequency of the metal,
$n$ - three-dimensional electron density, $m$ is the effective mass
of an electron.
The low frequency case is considered in the majority of the works.
Some works, however (see, say Ref.\cite{mar,stef}), devoted to the
high frequency case.

One of the most perspective methods of fabrication of the MPC
is the infiltrating various metals and semi-metals into the opals 
using high pressure\cite{zak1,zak2}. In our numerical
estimates below we mostly keep in mind these materials.

In this work we consider a two dimensional MPC consisting of metallic
circular cylinders in $z$ direction, embedded into
a dielectric media with the dielectric constant $\epsilon_m$.
The axes of the cylinders form a square lattice 
in the $x$-$y$ plane.
We assume a high frequency case so that dielectric constant
of a metal is given by Eq.~(\ref{ep}).
 
The important applications of MPC are connected with the idea
of creation a ``transparent metal'', the material, which has
a metallic type conductivity and at the same time is transparent
at infrared or visible range.
A regular bulk metal has the plasma frequency of the order of
a few electron volts and it completely reflects light of the lower
frequency.
We show here that our system reminds a transparent metal even for
the s-polarization (electric field is in $z$ direction).
It has a very narrow gap connected to a surface plasma wave and
it is transparent at all other frequencies.
 
A similar system has been considered theoretically 
by Kuzmiak et. al.\cite{mar}.
They assumed that the wave vector of EMW is perpendicular to the axes
of the cylinders. Their computations show that for s-polarization
the system completely reflects EMW with the frequency below
$\omega_p \sqrt{f/\epsilon_m} $, where $f$ is the volume fraction of
the metal in the system. We confirm this result but we show that
for the arbitrary angle of incidence the transparency
is non zero at any low frequency.  
An important advantage of our method is that it is pure analytical.  
This gives us a deeper insight in the physics of the wave propagation, 
which is difficult for computational methods.  Our method based
upon small parameters $f \ll 1$ and $kR \ll 1$, where $R$ is the radius of the cylinders. 
For the case considered by Kuzmiak et. al.\cite{mar} our result 
for s-polarization can be obtained
by a regular expansion of the system of equations derived in 
Ref.\cite{mar} with respect to parameters mentioned above.
 
The main feature of our physical picture is a pretty complicated
spectrum of the plasmons in the system, $\Omega_0({\bf k})$.
Now we consider a few limiting cases to get a qualitative idea about
this function.
Generically this spectrum comes from the one dimensional
plasmon. The spectrum of the one dimensional plasmon for one cylinder
only has a form~\cite{sh,kr}
\begin{equation} 
\label{oned} 
\omega^2(k_z) = \frac{\omega_p^2 R^2 k_z^2}{2 \epsilon_m}  
	\log{\frac{2}{k_z R}},
\end{equation} 
where $R$ is the radius of the cylinder.
This spectrum is almost linear and it starts from zero frequency.
The electric field of the plasmon outside the cylinder decays
exponentially as $\exp(-k_z \rho)$, where $\rho$ is the
cylindrical coordinate.
 
The physics is more sophisticated for 
a set of parallel cylinders which form a square
lattice in the plane ($x, y$) with a lattice constant $d$.
If the plasma wave is homogeneous in the ($x, y$) plane so that
$k_x = k_y = 0$, and $k_z d \ll 1$ then electrons in a given cylinder
feel electric field of many other neighboring cylinders.
Then electrons in all cylinders vibrate with the same
phase.
Such a mode is equivalent to a three dimensional plasma wave
with a frequency $\omega_p \sqrt{f/\epsilon_m}$.
If $k_zd \gg 1$ there is no interaction between
cylinders and one gets the one dimensional plasma wave.
However its spectrum starts with the frequency 
$\omega_p \sqrt{f/\epsilon_m}$
and there are no modes with a lower frequency.
 
On the other hand, if $k_x = k_y = \pi/d$, the phases
of 1D plasmons in the neighboring cylinders are opposite, the electric
fields created by all cylinders at a given one cancel each other.
In this case the plasma waves in the system are still one dimensional
and its spectrum starts with zero frequency.

Plasma wave with a
frequency $\Omega_0({\bf k})$
can be excited by EMW which has a non zero $E_z$ component.
This excitation is resonant if $\omega$ of the EMW
coincide with the corresponding value of plasma wave.
Note, that the bulk plasma wave can not be excited by EMW because the
first
one is pure longitudinal while the latter is pure transverse.
The plasma waves in our system are generated by electrons
moving in the cylinders only, while the EMW propagates everywhere
outside and inside the cylinders. Therefore in this case the 
EMW does excite plasma
modes.

There is another plasma mode in our system, which can be excited
by electric field with non zero component perpendicular 
to the cylinders.
This is a surface plasma mode with a frequency 
$\omega_s = \omega_p/\sqrt{\epsilon_m + 1}$.
This mode also plays the important role in the propagation
of EMW.

In this paper we show that there are five different branches of the EMW
which cover all the frequency range with $\omega \tau \gg 1$ with
one omnidirectional gap in the vicinity of the $\omega_s$.
Except of this gap 
there is another gap at lower frequencies which is not omnidirectional
and which depends on the polarization of the EMW.
The gap obtained by Kuzmiak et. al.~\cite{mar}
is a particular case of such gap at $k_z = 0$ and $s$-polarization
of EMW.

Our paper is organized as follows.
In section \ref{eigenfreq} we derive the spectrum of the plasma waves in the system.
In section \ref{secthree} we study the dielectric properties of the system.
They are described by the tensor of dielectric constants.
In the principal axes ($x, y, z$) two components
of the tensor are equal to each
other, namely $\epsilon_{xx} = \epsilon_{yy} = \epsilon_{\perp}$ and they
depend on the frequency only.
The longitudinal component
$\epsilon_{zz} = \epsilon_{\|}$ is a function of $\omega$ and ${\bf
k}$.
The non-locality reflects the extra degrees of freedom and the 
long range Coulomb interaction.
To calculate $\epsilon_{\|}$ one should take into account
deviation of acting field from average field.
In section \ref{secfour} we consider EMW in the system. They may be classified
as ordinary and extraordinary waves by analogy with
the optics of uniaxial dielectric crystals.
However since the dielectric tensor is a function of $\omega$
and ${\bf k}$ we get two branches for the ordinary wave and
three branches for the extraordinary one.
In section \ref{secfive} reflection and refraction of EMW is
considered.
As in the case of optics of uniaxial crystals we obtain double
refraction of EMW.

\section{Spectrum of the plasma waves} 
\label{eigenfreq} 
 
The two-dimensional model considered in this paper consists
of infinite metallic cylinders of a circular
cross-section.
The cylinders are parallel to the $z$ axis and form a square
lattice with the
lattice constant $d$.
Space between the cylinders is filled with a dielectric with the dielectric
constant $\epsilon_m$.

To calculate plasma waves one should assume
that the three-dimensional electron density
in the $l^{\rm th}$ cylinder with coordinates $(x_l,\ y_l)$ has a modulation
\begin{equation} 
n_l = n + n' \cos(k_z z)\, e^{i(k_x x_l + k_y y_l - \omega t)}.
\end{equation} 
In this case electrostatic potential from the $l^{\rm th}$ cylinder
at a point with coordinates $(x, y, z)$ has a form

\begin{equation} 
\Phi_l(x, y, z) = A \cos(k_z z) 
	e^{i(k_x x_l + k_y y_l - \omega t)}
	K_0\left(k_z \sqrt{(x-x_l)^2 + (y-y_l)^2}\right),
\end{equation} 
where $A=2 \pi n' e R^2/\epsilon_m$, 
$K_0$ - modified Bessel function of $0$-th order.
 
The potential at the point $(x, y, z)$
created by the whole system of cylinders can be written as
\begin{equation} 
\label{potential} 
\Phi(x, y, z) = \sum_{all\ wires} \Phi_l(x,\ y,\ z) = A \cos(k_z z)	
	e^{i(k_x x + k_y y - \omega t)} \varphi(x, y),\end{equation} 
where
\begin{equation} 
\label{phi} 
\varphi (x, y) = \sum_{n_x,n_y=-\infty} 
	^{\infty} e^{-i (k_x (x - n_x d) + k_y (y - n_y d))}
	K_0\left( k_z \sqrt{(x-n_x d)^2+(y-n_y d)^2 + R^2}\right).
\end{equation} 
Since $R \ll d$ the term $R^2$ under the square root is important 
for $x - n_x d = y - n_y d = 0$ only.
Note that $\varphi (x, y)$ is a periodic function with the
period $d$ with respect to both arguments.

Equation of motion for electrons in a cylinder is
\begin{equation} 
\label{emotion} 
m \frac{dV_z}{dt} = -e\frac{\partial \Phi}{\partial z},
\end{equation} 
where $V_z$ is a drift velocity, so that the current density
$j_z = e n V_z$.
The continuity equation is
\begin{equation} 
\label{cont} 
e \frac{\partial n'}{\partial z} + \frac{d j_z}{d z} = 0.
\end{equation} 
Using Eqs.~(\ref{potential}, \ref{emotion}, \ref{cont})
one can obtain
the plasma frequency of the system
\begin{equation} 
\label{plasmafreq} 
\Omega_0^2(k_x, k_y, k_z) = \omega_{k_0}^2\, \varphi_0(k_x, k_y, k_z), 
\end{equation} 
where
\begin{equation} 
\omega_{k_0}^2 = \frac{f \omega_p^2 d\,^2}{2 \pi \epsilon_m} k_z^2, 
\end{equation} 
\begin{equation} 
\label{phi0} 
\varphi_0(k_x, k_y, k_z) = \mathop{{\sum}'}_{n_x,n_y=-\infty} 
	^{\infty} e^{i(k_x n_x + k_y n_y)d}\, K_0\left(k_z d
	\sqrt{n_x^2+n_y^2}\right) + K_0(k_z R).
\end{equation} 
Here $\mathop{{\sum}'}$ means that we should exclude the term
$n_x = n_y = 0$ from the summation.
One can see that the function $\Omega_0$ is a periodic function
of $k_x$ and $k_y$ with the periods equal to reciprocal vectors of
the square lattice, but it is not periodic with respect to $k_z$.
 
In the case when $k_x = k_y = 0$ and $k_z d \ll 1$ we obtain that
$\varphi_0(0, 0, k_z) = 2 \pi/k_z^2 d^2$, thus the plasma frequency of the
system
$\Omega_0^2(0, 0, k_z) =\omega_p^2  (f/\epsilon_m)$.
In another limiting case $k_x = k_y = 0$ and $k_z d \gg 1$ the
plasma frequency is given by Eq.~(\ref{oned})
for the one dimensional plasmon.
The same result is valid at any $k_z$ if $k_x = k_y = \pi/d$.
 
\begin{figure}
\centering\epsfig{file=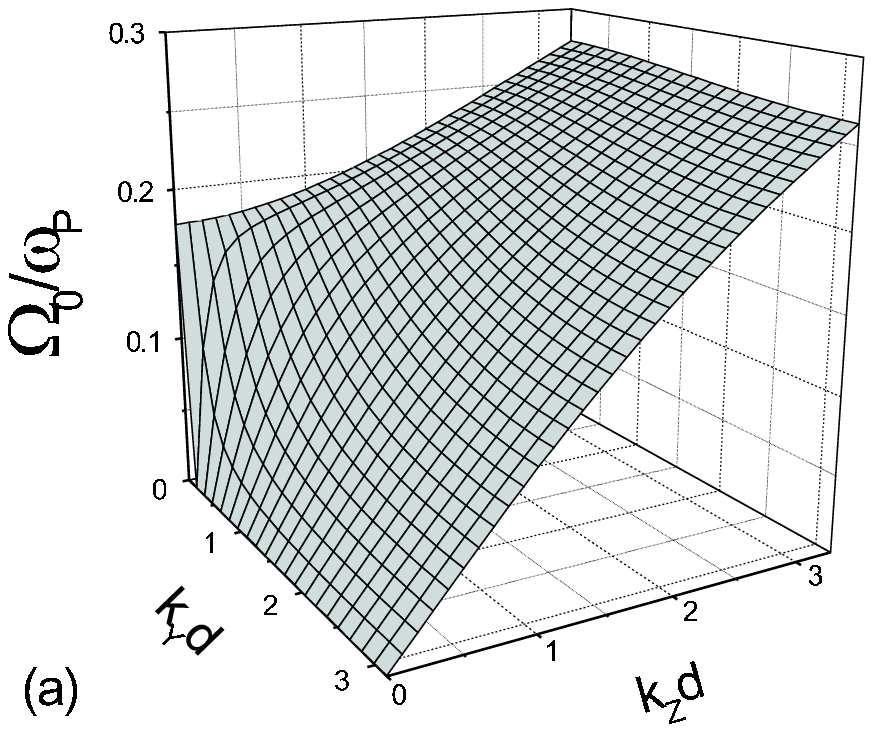, width=12cm}
\centering\epsfig{file=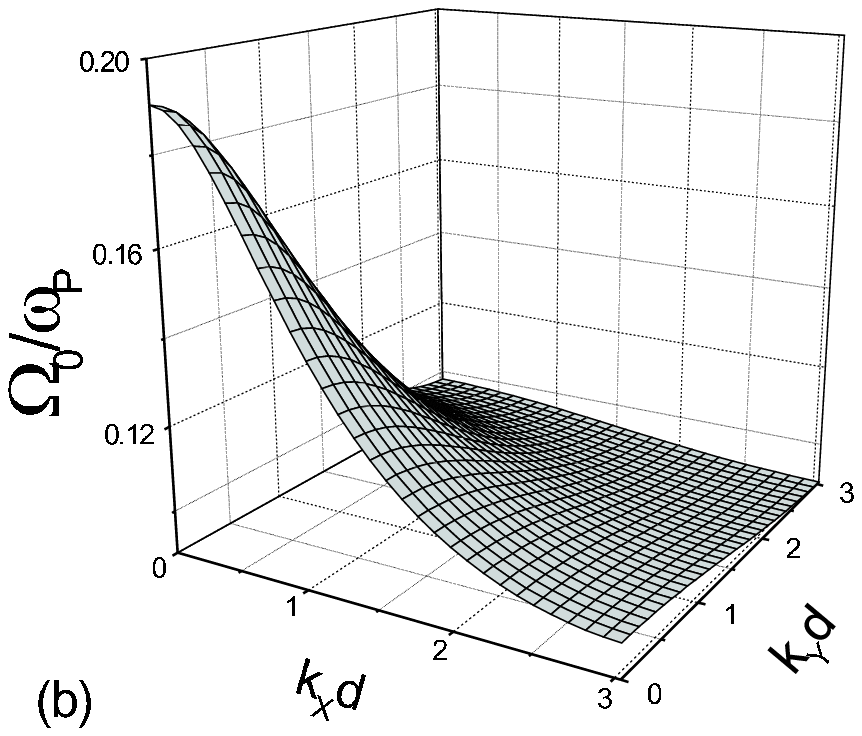, width=12cm}
\vspace{0.5cm}
\caption{Plasma frequency $\Omega_0({\bf k})$ in the plane $k_x = 0$ (a) and 
 	in the plane $k_z d = 0.5$ within one quarter of the 
	first Brillouin zone (b) at the following values of parameters 
	$f = 0.1$, $c/\omega_p d = 0.03$. }
\label{omega3d}
\end{figure}

Figs. \ref{omega3d}(a) and \ref{omega3d}(b) illustrate the behavior of the
$\Omega_0(k_x, k_y, k_z)$ as a function of $k_x, k_y, k_z$.

\section{Dielectric properties of the system} 
\label{secthree}

\subsection{ Calculation of the $\epsilon_{zz} = \epsilon_{\|}$ 
	component of the dielectric tensor}

Let us introduce an external electric field acting in $z$ direction in
a form
\begin{equation} 
{\cal E}_{\rm ext}= {\cal E}'_{\rm ext} e^{i(k_x x + k_y y + k_z z
-\omega t)}.
\end{equation}

Now the equation of motion~(\ref{emotion}) has a form
\begin{equation} 
	m\frac{dV_z}{dt} = e({\cal E}_{\rm ext} + {\cal E}).
\end{equation} 
Here ${\cal E}$ is an internal field in $z$ direction given by
\begin{equation} 
\label{elfield}
{\cal E}(x, y) = -\frac{\partial \Phi}{\partial z}
	= -i k_z \frac{2 \pi e R^2}{\epsilon_m} n'\,
	e^{i(k_x x + k_y y + k_z z - \omega t)} \varphi(x, y).
\end{equation} 
Using the continuity equation~(\ref{cont}) and the results of
Sec.~\ref{eigenfreq} one gets\begin{equation} 
\label{n} 
n' = \frac{i e n k_z}
	{m\left(\omega^2 - \Omega_0^2(k_x, k_y, k_z)\right)}{\cal
	E}'_{ext}.
\end{equation}

Now we can relate current density $j_z$ to the external field
\begin{equation} 
\label{current} 
j_z = \frac{\omega n'}{k_z} = \frac{i e n \omega}
	{m\left(\omega^2 - \Omega_0^2(k_x, k_y, k_z)\right)}{\cal E}'_{ext}.
\end{equation} 
 
To find effective macroscopic conductivity $\sigma$ 
one should relate average current density $\overline{\jmath}_z$ to
the average electric field $E$ by equation
\begin{equation}
\label{ohm}
\overline{\jmath}_z = \sigma E.
\end{equation}
The field $E$ can be found by the direct averaging
\begin{equation} 
\label{eaverage} 
E = {\cal E}_{\rm ext} + \frac{1}{d\,^2}
	\int\!\!\int_{\rm cell}\ {\cal E}(x,y)\ dx dy,
\end{equation} 
while 
\begin{equation}
\label{jave}
\overline{\jmath}_z = f j_z.
\end{equation}
Using equations ~(\ref{elfield})~and~(\ref{n})
one can obtain a relation
between the average and the external fields
\begin{equation} 
\label{extandave} 
E = {\cal E}_{ext}\left(1 + \frac{\omega_{k_0}^2
	\overline{\varphi}(k_x, k_y, k_z)}
	{\omega^2 - \Omega_0^2(k_x, k_y, k_z)}\right),
\end{equation}  
where
\begin{equation} 
\label{phiave} 
\overline{\varphi}(k_x, k_y, k_z) = \frac{1}{d\,^2} 
	\int\!\!\int_{\rm cell}\ \varphi(x, y)\ dxdy
\end{equation} 
After double summation and integration in Eq.~(\ref{phiave}) one can
obtain
\begin{equation} 
\overline{\varphi}(k_x, k_y, k_z) = \frac{2 \pi}{(k_x^2 + k_y^2 + k_z^2)d\,^2}. 
\end{equation} 
 
Using Eqs.(\ref{current}),(\ref{ohm}),(\ref{jave}),(\ref{extandave}) 
one can find $\sigma$.
Since the dielectric constant
\begin{equation} 
\epsilon = \epsilon_m + i\frac{4 \pi \sigma}{\omega} 
\end{equation} 
one can obtain for the $\epsilon_{zz} = \epsilon_{\|}$
component of the dielectric tensor
\begin{equation} 
\label{epspar} 
\epsilon_{\|}(k_x, k_y, k_z, \omega) = \epsilon_m - \frac{f \omega_p^2} 
	{\omega^2 - \Omega^2(k_x, k_y, k_z)},
\end{equation} 
where
\begin{equation} 
\Omega^2(k_x, k_y, k_z) = \Omega_0^2(k_x, k_y, k_z) -  
	\frac{f}{\epsilon_m}\omega_p^2\frac{k_z^2}{k_x^2+k_y^2+k_z^2}.
\end{equation} 
 
Until now we assumed that relaxation time of the electrons $\tau$ is
infinite.
To describe the decay of the EMW one should substitute $\omega^2$ by
$\omega(\omega + i \tau^{-1})$ in Eq.~(\ref{epspar}).

\begin{figure} 
\centering\epsfig{file=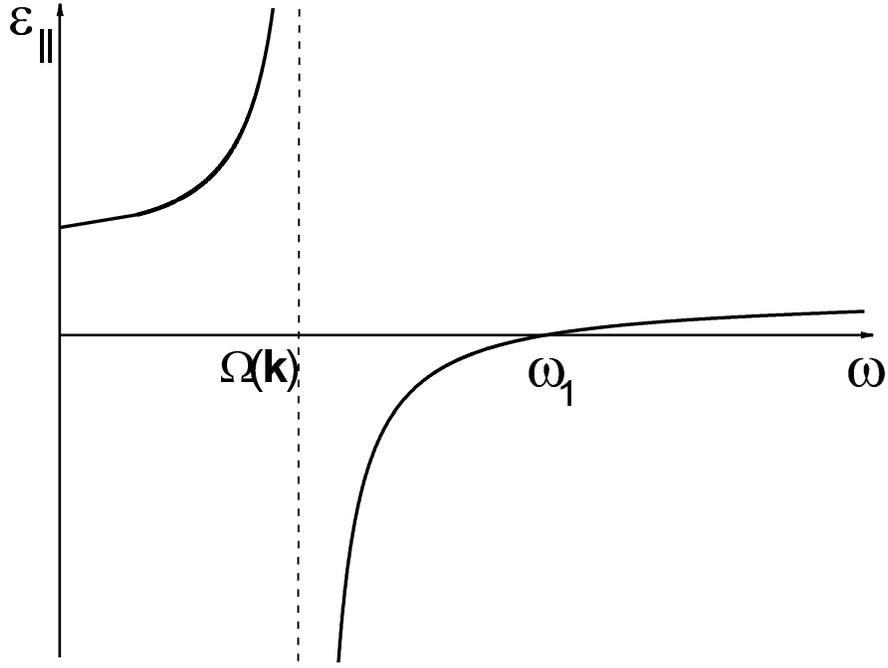, width=12cm}
\vspace{0.5cm}
\caption{Dielectric constant $\epsilon_{\|}$ as a function of
  	frequency at a fixed value of {\bf k} with non zero $k_z$. 
	Relaxation time is taken to be infinite.}
\label{epsom}
\end{figure}

Dependence of $\epsilon_{\|}$ on the frequency
at a fixed value of the wave vector ${\bf k}$ is shown at
the Fig.~\ref{epsom} for the case $k_z \neq 0$.
One can see that at small frequency $\epsilon_{\|}$ is positive.
The dielectric constant $\epsilon_{\|}$ becomes zero at
$\omega = \omega_1 = (\Omega^2 + f \omega_p^2)^{1/2}$.
Note that the region of the negative $\epsilon_{\|}$ 
is not omnidirectional and it disappears at some values of ${\bf k}$.

\subsection{Calculation of the  
	$\epsilon_{xx} = \epsilon_{yy} = \epsilon_{\perp}$
	components of the dielectric tensor}
 
Assuming that $f \ll 1$ one can apply the method for finding the dielectric
constant of the granular mixture, described by Landau and Lifshitz
(see Ref.\cite{land}, p. 45).
One starts from the identity
\begin{equation} 
{1\over V}\int ({\bf {\cal D}(r)}-\epsilon_m {\bf {\cal E}(r))}dV={\bf D}-\epsilon_m {\bf E},
\label{landau} 
\end{equation} 
where ${\bf D}$ and ${\bf E}$ are average induction and electric field
respectively, $V$ is the volume of the system.
In the following calculations we assume that all fields are perpendicular
to the cylinders.
The dielectric constant is determined
by relation ${\bf D}=\epsilon_{\perp} {\bf E}$.
The integrand in the left side of Eq.~(\ref{landau}) is non-zero only
inside the cylinders.
Since $f$ is small one can assume that electric field
acting on a cylinder is ${\bf E}$.
The field inside the cylinder ${\bf E}_{in}$ is constant and equal
(see Ref.\cite{land}, p. 43 )
\begin{equation} 
{\bf E}_{in}={2\epsilon_m \over \epsilon_{in} + \epsilon_m}{\bf E},
\end{equation} 
where internal dielectric constant $\epsilon_{in}$ is
given by~Eq.(\ref{ep}).
As the result the component of the dielectric tensor
$\epsilon_{\perp}$ is
\begin{equation} 
\epsilon_{\perp}(\omega) = \epsilon_m - 2 f \epsilon_m {(\epsilon_m-1) \omega^2  
	+ \omega_p^2
	\over (\epsilon_m+1) \omega^2 - \omega_p^2}.
\label{eperp} 
\end{equation} 
Note that the pole in Eq.~(\ref{eperp}) corresponds to a surface plasma
wave propagating along the surface of the cylinder.
 
This equation is written in the first approximation with respect to the
small parameter~$f$.
This approximation can be improved by a
two-dimensional version of Clausius-Mossotti procedure.
One can show that effective field is ${\bf E}+2\pi {\bf P}$,
where ${\bf P}$ is the polarization.
As a result one gets an extra term $2 f\omega_p^2$
in the denominator of Eq.(\ref{eperp}) which
can be neglected.

Finally, we have found all diagonal components of the
dielectric tensor $\epsilon_{zz} = \epsilon_{\|}$ and
$\epsilon_{xx} = \epsilon_{yy} = \epsilon_{\perp}$.
Since $z$ is chosen in the direction of the axes of the cylinders
the dielectric tensor is diagonal at any choice of $x$ and $y$.

\section{Spectrum of electromagnetic waves in the system} 
\label{secfour} 
 
The Maxwell's equation for the EMW has a form (see Ref.\cite{land}, p. 316)
\begin{equation}
\label{max}
	n^2 E_i - n_i (n_k E_k) = \epsilon_{ik} E_k,
\end{equation}
where
\begin{equation} 
{\bf n} = \frac{c}{\omega}\, {\bf k}.
\end{equation}
The spectrum of EMW can be found from the equation
\begin{equation} \label{det} 
{\rm det} | n^2 \delta_{ik} - n_i n_k - \epsilon_{ik}(n_x, n_y, n_z) |= 0.
\end{equation} 
 
Since the structure of the dielectric tensor is the same as for a uniaxial
crystal one can get that Eq.(\ref{det}) falls off into
two equations:
\begin{equation} 
\label{ord} 
n^2 = \epsilon_{\perp}(\omega)
\end{equation} 
and
\begin{equation} 
\label{exord} 
\frac{1}{n^2}=\frac{\sin^2\theta}{\epsilon_{\|}(n_x, n_y, n_z, \omega)}  
	+ \frac{\cos^2\theta}{\epsilon_{\perp}(\omega)}
\end{equation} 
where $\theta$ is the angle between wave vector ${\bf k}$ and $z$ axis.
Using analogy with the uniaxial dielectric crystals
we call 
the solutions of
Eq.(\ref{ord}) and Eq.(\ref{exord}) the ordinary 
and the extraordinary waves respectively.
In the ordinary waves $E_z$ component of the electric field is zero.

In fact, our equations are much more difficult than in the
regular optics of dielectric crystals because 
of the complicated dependences
of the dielectric tensor on $\omega$ and ${\bf k}$.
Due to these dependences there are two ordinary waves and three
extraordinary waves.
 
\begin{figure}
\centering\epsfig{file=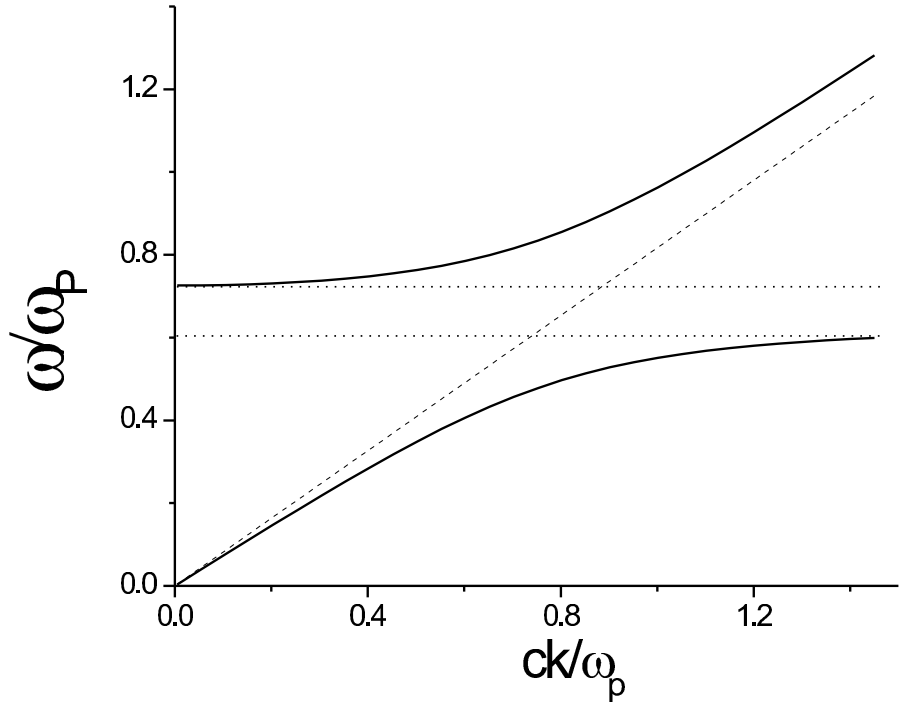, width=12cm}
\vspace{0.5cm}  
\caption{The spectra of the ordinary wave as given by
	Eq.(\ref{solord}) (solid lines).
	The lower dotted horizontal line is
	$\omega = \omega_s$. The upper dotted line is
	$\omega = \omega_s (1+f\epsilon_m)$.
	The dashed line shows the photon dispersion
	$\omega = c k /\epsilon_m(1-f)$.
	Note that the solid lines represent also the spectrum
	of the extraordinary waves at $\theta = 0$.  $f = 0.1$,	$c/\omega_p d = 0.03$}
\label{dspord}
\end{figure}

The frequencies of the two ordinary waves depend only on $|k|$ and
can be obtained analytically
$$
\omega_o^2 = \frac{\epsilon_m (1+2f)\omega_p^2 + c^2 k^2 (\epsilon_m +1)}
	{2\epsilon_m (\epsilon_m (1-2f) + 1 + 2f)}
$$
\begin{equation}
\label{solord}  
	\qquad {}
	\pm
	\frac{\sqrt{\left( \epsilon_m (1+2f)\omega_p^2 + c^2 k^2	
	(\epsilon_m +1) \right)^2
	- 4 c^2 k^2 \omega_p^2 \epsilon_m (\epsilon_m (1-2f) + 1 +
          2f)}}
	{2\epsilon_m (\epsilon_m (1-2f) + 1 + 2f)}
\end{equation}  
These solutions are shown in the Fig.~\ref{dspord}.
All the calculations shown in the plots below are made for
$f = 0.1$ and $c/\omega_p d = 0.03$.
The lower branch of the ordinary wave
tends to the frequency of the surface plasmon mode
$\omega_s = \omega_p/\sqrt{\epsilon_m + 1}$ as $c k/\omega_p$
tends to infinity.
The upper branch starts with the frequency $\omega_s (1+f\epsilon_m)$and
transforms into a regular dispersion law for the EMW in the system
with the dielectric constant $\epsilon_m(1-f)$.
The spectrum of the ordinary wave has a gap of the width
$f \epsilon_m \omega_s$. As we show below, a part of this
gap is covered by the extraordinary waves, however a small
omnidirectional
band gap still exist in this region.

To find a dispersion relation for the extraordinary wave one
should solve cubic equation for $\omega^2$ (Eq.~(\ref{exord})).
Since the analytical solution is not instructive
we present only the results of numerical calculations.
The component of the dielectric tensor $\epsilon_{\|}$ depends
on $|k|, \theta$, and the azimuthal angle $\phi$.

\begin{figure}
\centering\epsfig{file=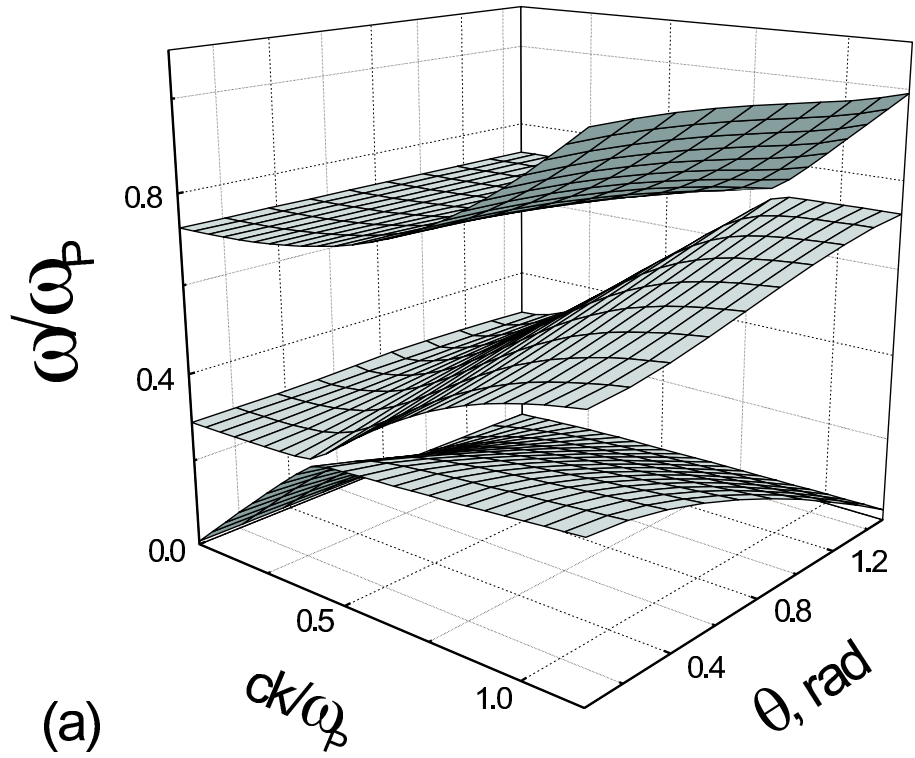, width=12cm}
\centering\epsfig{file=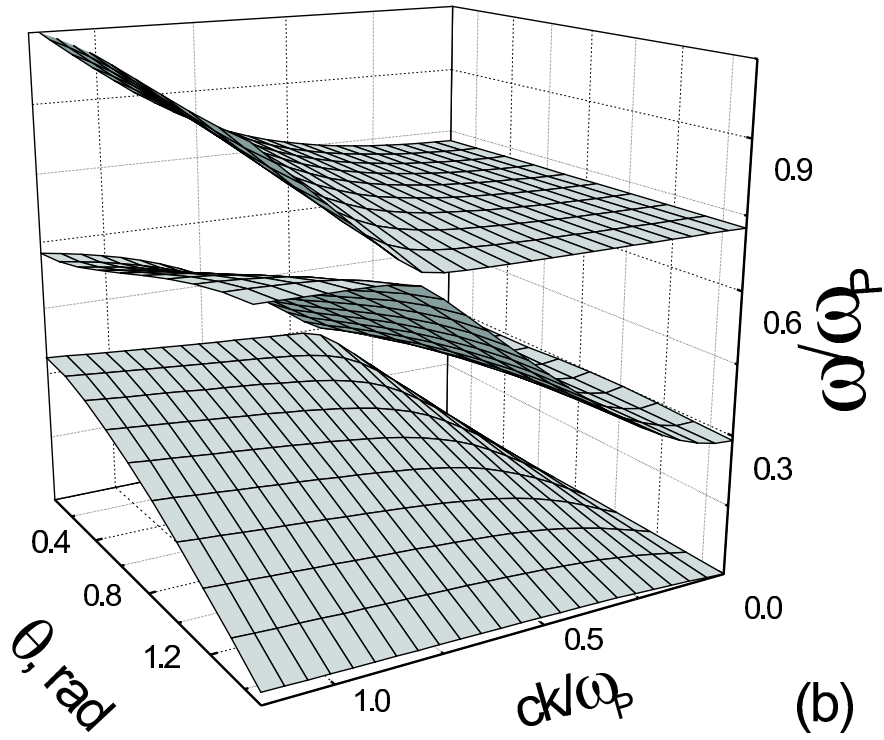, width=12cm}
\vspace{0.5cm}
\caption{Three branches of
	extraordinary waves in coordinates $c k/\omega_p$ and $\theta$
	from different points of view.  $f = 0.1$,
	$c/\omega_p d = 0.03$.}
\label{dsp3d}
\end{figure}

\begin{figure} 
\centering\epsfig{file=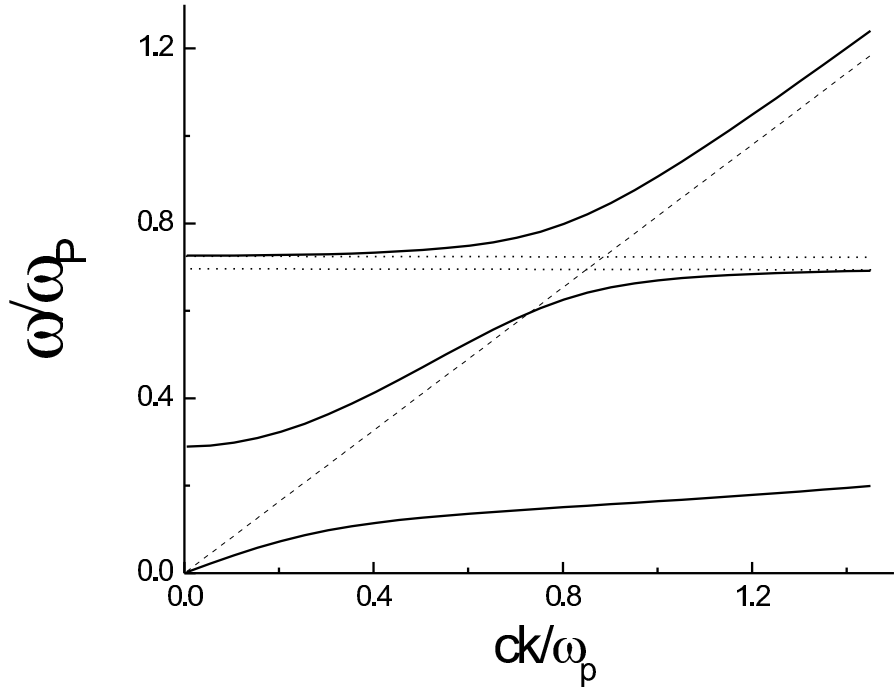, width=12cm}
\vspace{0.5cm}  
\caption{Three extraordinary branches at $\theta = 57^{\circ}, \phi  = 45^{\circ}$.
	The lower dotted horizontal line is
	$\omega = \omega_s (1 +f \sin^2\theta (\epsilon_m - 1)/(\epsilon_m + 1))$. 
	The upper dotted line is
	$\omega = \omega_s (1+f\epsilon_m)$.
	The dashed line shows the photon dispersion
	$\omega = c k /\epsilon_m(1-f)$.
	$f = 0.1$, $c/\omega_p d = 0.03$. }
\label{dspexord}
\end{figure}

\begin{figure} 
\centering\epsfig{file=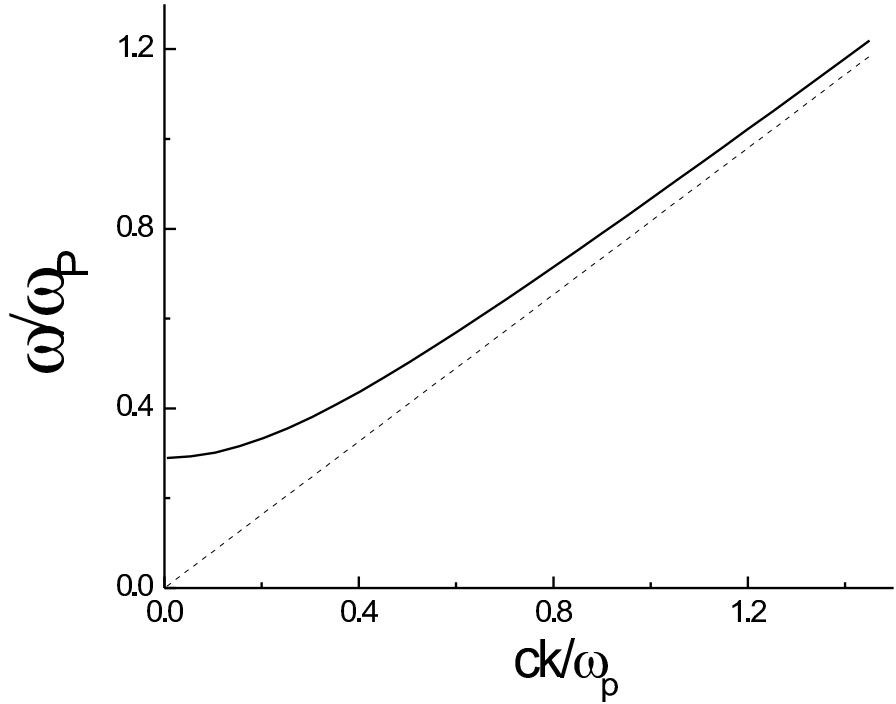, width=12cm}
\vspace{0.5cm}  
\caption{The only extraordinary branch at $\theta = \pi/2$. 
	The dashed line shows the photon dispersion
	$\omega = c k /\epsilon_m(1-f)$.
	 $f = 0.1$, $c/\omega_p d = 0.03$. }
\label{thPi2}
\end{figure}

Now we demonstrate the results of our calculations 
of the EMW spectra in the first Brillouin zone only. 
The EMW spectra in higher zones can be easily obtained
but since we keep in mind small values of $d$ they correspond
to pretty high frequencies.
The 3D plots Fig.~\ref{dsp3d}(a) and (b) show three branches of
extraordinary waves in coordinates $c k/\omega_p$ and $\theta$
from different points of view.
The azimuthal angle $\phi$ is chosen $45^{\circ}$ for all plots below.
Fig.~\ref{dspexord} shows three extraordinary branches
at $\theta = 57^{\circ}, \phi = 45^{\circ}$.
Fig.~\ref{dspord} shows extraordinary branches at $\theta = 0$.
One can see from Eq.~(\ref{exord}) that at $\theta = 0$ the extraordinary
branches coincide with the ordinary ones.
Fig.~\ref{thPi2} shows extraordinary wave at $\theta = \pi/2$.
Note that there are singularities at $\theta = 0$ and $\theta = \pi/2$
(see Eq.~(\ref{exord})),
because at $\theta = 0$ the zero of $\epsilon_{\|}$ does not
play any role.
The same happens with the zero of $\epsilon_{\perp}$ at $\theta =
\pi/2$.
Therefore Fig.~\ref{dspord} has only two branches.
The lower extraordinary branch in Fig.~\ref{dspord}
is the result of unification of the lower and the middle branches
of the  Fig.~\ref{dsp3d}(a).
They form one branch which coincide 
with the lower branch at small $k$ and
with the middle branch at large $k$.
The same happens with the middle and upper branches
in Fig.~\ref{dsp3d}(b).
They form one branch which coincide with the middle branch at small $k$
and with the upper branch at large $k$.
Since the lower branch is absent at $\theta = \pi/2$, 
Fig.~\ref{thPi2} has only one branch which coincides with
the results of the Ref.\cite{mar}.
Note, that all singularities will be
smeared out if one takes into account the finite value of $\tau$.

The lower branch in Fig.~\ref{dspexord} starts with the 
linear dispersion law
$\omega = c k \cos\theta/\sqrt{\epsilon_m}$ .
At $k_z \gg 1/d$ this branch tends to the one dimensional plasmon
with a dispersion law Eq.(\ref{oned}),
but its frequency remains higher than the frequency of the
plasmon.

The middle branch starts with the frequency
$\omega_p (f/\epsilon_m)^{1/2}$ and saturates at
$\omega_s (1 +f \sin^2\theta (\epsilon_m - 1)/(\epsilon_m + 1))$,
which
is above saturation value of the lower branch of the ordinary wave
$\omega_s$.
The upper branch starts at the $\omega_s (1+f\epsilon_m)$.
Note, that the upper branch of the ordinary wave starts with
the same frequency.
It is easy to show that the omnidirectional gap
is very narrow and equal to
$f \omega_s (\epsilon_m^2 + 1)/(\epsilon_m+1)$.
At all frequencies below and above this gap the system is
transparent.
The upper branch transforms into the regular photon with the
dispersion $c k /\epsilon_m(1-f)$.

\section{Reflection and refraction of electromagnetic wave} 
\label{secfive}

\begin{figure}
\centering\epsfig{file=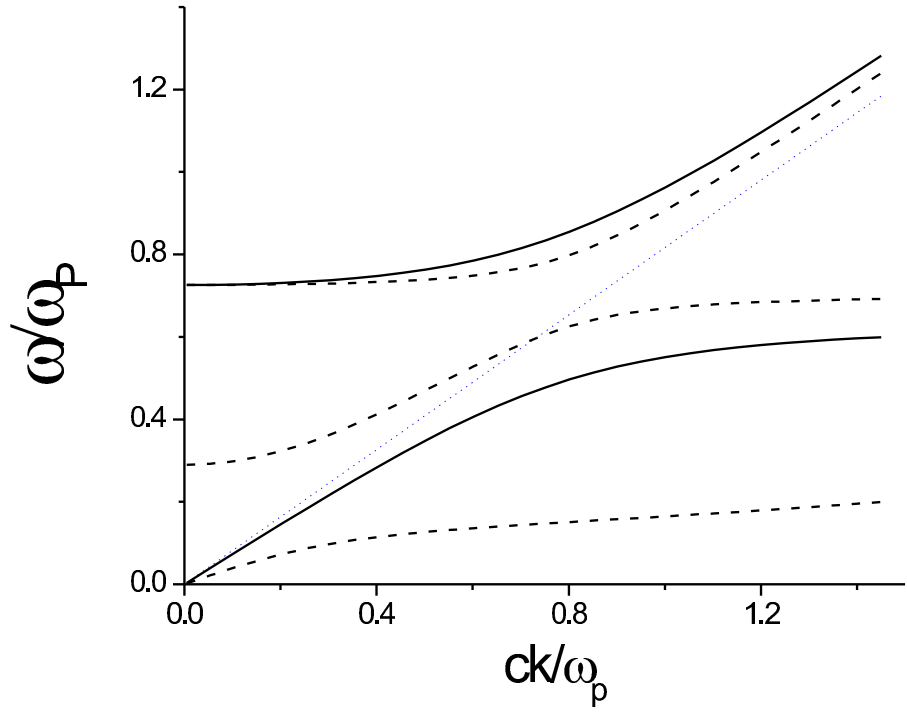, width=12cm}
\vspace{0.5cm} 
\caption{Dispersion of ordinary (full line) and extraordinary wave 
	(dashed line). This figure combines Fig.~\ref{dspord} and 
	Fig.~\ref{dspexord}. The dotted line shows the photon dispersion
	$\omega = c k /\epsilon_m(1-f)$.}
\label{dspboth}
\end{figure}

Fig.~\ref{dspboth} shows complete spectrum of electromagnetic waves
in the
system, which includes both ordinary and extraordinary waves.
One can see, that every frequency corresponds to two waves
(ordinary and extraordinary) everywhere except the gap for the ordinary waves.
Thus, in a general case, one should expect a 
double refraction similar to a regular
optics of dielectric crystals.
Our case is more difficult because of the $\omega$ and ${\bf k}$
dependence of the dielectric constant.

We mention first that tangential component of the wave vector of
incident, reflected and refracted waves should be the same.  
This follows from the very general statement (see Ref.~\cite{jackson}, p. 304) 
that the
spatial variation and time variation of the electric and magnetic field 
vectors must be the same at the boundary.  
Suppose that the
boundary is the plane $\xi = 0$ and the plane of the incidence is ($\xi, \eta$).  
Then one gets
\begin{equation} 
n_{r\eta} = n_{1\eta} = n_{2\eta} = n_{i\eta} \equiv \sin\theta_i\, ,
\end{equation} 
where symbols $r$, $1$, $2$, $i$ label reflected, ordinary refracted,
extraordinary refracted and incident wave respectively, 
$\theta_i$ is the angle of incidence.
 
The solution of the problem of reflection and refraction in the case
of ${\bf k}$ dependent dielectric constant should consist of the
following parts.  
The first part is to find $n_{1\xi}$ and $n_{2\xi}$
which determine angles of refraction of ordinary and extraordinary 
waves $\theta_1$ and $\theta_2$ respectively.
For the ordinary wave refraction angle can be found using the Snell's
law
\begin{equation} 
	\sin\theta_1 = \frac{\sin\theta_i}{\sqrt{\epsilon_{\perp}}}. 
\end{equation} 
Here and below we assume that incident and reflected waves propagate
in the vacuum.
 
Due to spatial dispersion of the $\epsilon_{\|}$ calculation of the
refraction angle for extraordinary wave is more complicated.  One
should find a solution for the frequency of the proper branch of the
extraordinary wave in a form $\omega = f({\bf n}_2)$.  This equation
can be used to find $n_{2\xi}$ at given $n_\eta$ and $\omega$.  The
refraction angle for the extraordinary wave can be found using
relation $\tan\theta_2 = n_\eta/n_{2\xi}$.

The second part of the problem is to find electric field of reflected
${\bf E}_r = (E_{r\eta}, E_{r\zeta}, E_{r\xi})$, ordinary refracted 
${\bf E}_1 = (E_{1\eta}, E_{1\zeta}, E_{1\xi})$ and extraordinary refracted 
${\bf E}_2 = (E_{2\eta}, E_{2\zeta}, E_{2\xi})$ waves.  
For this purpose we should
first find two relations between three components of vector ${\bf E}$
in each refracted wave from Eq.~(\ref{max}).  The other equations are
the regular boundary condition on the surface including the continuity
of the normal component of the displacement vector $D_i =
\epsilon_{ik} E_k$.

Strictly speaking the last condition should be formulated in
coordinate space rather than in ${\bf k}$ space \cite{agran}.  However it
works in ${\bf k}$ space if $kd \ll 1$.  In general case one can use
the continuity of the normal component of the Poynting vector,
calculated far from the boundary.  The scheme described above is valid
when the boundary plane forms any angle with the axis of cylinders in
our system.  Note that if electric field in the incident wave is
perpendicular to the cylinders, there is only ordinary
refracted wave.  In this case equations for all fields have a regular
form (see Ref.~\cite{jackson}, p. 305).

As an example we consider a case when the boundary plane is
perpendicular to the cylinders (axis $\xi$ coincides with the $z$
axis) and the plane of incidence includes $z$ axis and the diagonal of
the square lattice formed by the axes of the cylinders in the plane 
$z = 0$.
Thus for all waves $n_x = n_y = \sin\theta_i / \sqrt{2}$.  For the
ordinary wave $n_{1z}^2 = \epsilon_{\perp} - \sin^2\theta_i$.  Suppose that we
have found $n_{2z} (\omega)$ for the extraordinary wave.  Note that
in different frequency ranges $n_{2z}(\omega)$ is determined by
different branches of extraordinary waves.  Then for electric fields
of the reflected and refracted waves one gets
\begin{equation}
\label{erx}
E_{rx} = \left( \frac{\cos\theta_i}{\Lambda} - \frac{1}{n_{1z} + 1}\right)
	E_{ix} 
	+ \left( \frac{\cos\theta_i}{\Lambda} - \frac{n_{1z}}{n_{1z} + 1}
	\right) E_{iy},
\end{equation}
\begin{equation}
\label{ery}
E_{ry} = \left( \frac{\cos\theta_i}{\Lambda} - \frac{n_{1z}}{n_{1z} + 1}\right)E_{ix} 
	+ \left( \frac{\cos\theta_i}{\Lambda} - \frac{1}{n_{1z} + 1}\right) E_{iy},
\end{equation}
\begin{equation}
\label{erz}
E_{rz} = \left( 1 - 2\frac{\cos^2\theta_i \tan\theta_2}{\sin\theta_i} 
	\frac{\epsilon_{\perp}}{\Lambda}\right)E_{iz},
\end{equation}
\begin{equation}
\label{e1xyz}
E_{1x} = -E_{1y} = \frac{1}{n_{1z} + 1} (E_{ix} - E_{iy}), \qquad E_{1z} = 0,
\end{equation}
\begin{equation}
\label{e2xy}
E_{2x} = E_{2y} = \frac{\cos\theta_i}{\Lambda} (E_{ix} + E_{iy}),
\end{equation}
\begin{equation}
\label{e2z}
E_{2z} = 2\frac{\cos^2\theta_i \tan\theta_2}
	{\sin\theta_i} \frac{\epsilon_{\perp}}{\Lambda \epsilon_{\|}(n_x, n_y, n_{2z})}E_{iz},
\end{equation}
where
\begin{equation}
\Lambda = \tan\theta_2 \sin\theta_i \left( 
	\frac{\epsilon_{\perp}} {\epsilon_{\|}(n_x, n_y, n_{2z})}
	-\epsilon_{\perp}\right) + \cos\theta_i + n_{2z}
\end{equation}
Reflection coefficient ${\cal R}$ can be found using 
Eqs.~(\ref{erx},\ref{ery},\ref{erz}) as
\begin{equation}
{\cal R} = \frac{E_r}{E_i}.
\end{equation}

\begin{figure} 
\centering\epsfig{file=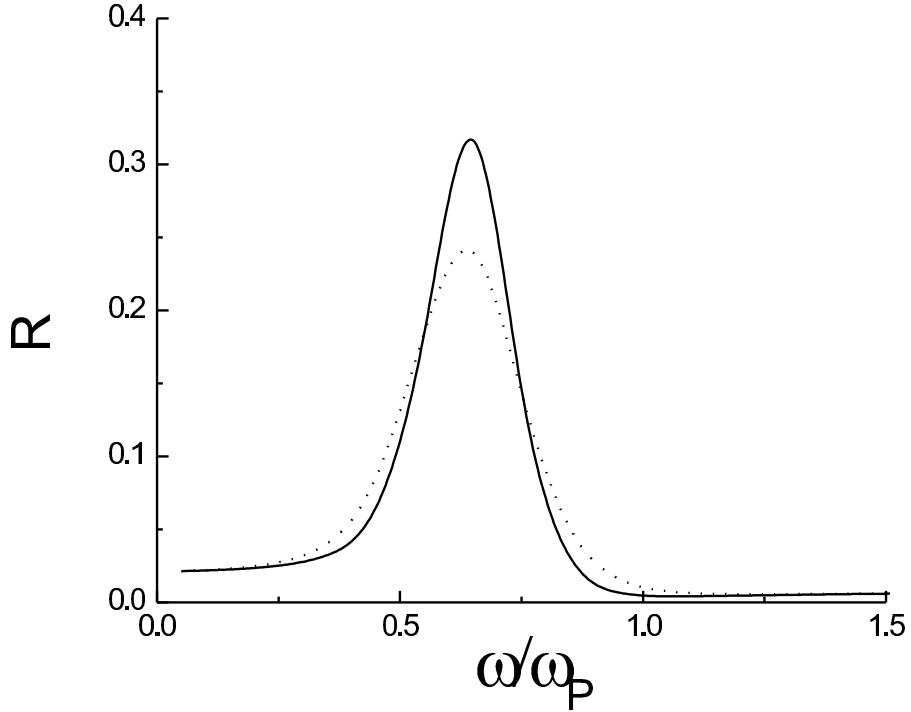, width=12cm}
\vspace{0.5cm}
\caption{Reflection coefficient for the case 
	of normal incidence 
	for the following values of parameters
	$\omega_p \tau = 10$ (solid line), $\omega_p \tau = 5$ (dotted line), 
	$\epsilon_m = 1.5$, $f = 0.1$.  }  
\label{refl}
\end{figure}

Fig.~\ref{refl} shows the 
reflection coefficient in the case of normal incidence.
For this case $E_z = 0$ in 
reflected and refracted
waves and only the ordinary wave propagates through the system.
To avoid singularities we assumed finite value of the relaxation time.
One can see that the system is basically transparent except for the small gap in the spectrum of EMW
in the vicinity of the surface plasmon.

\section{Conclusion}

We have found that the two dimensional metallic mesh has two different plasma
modes. One of them comes generically from one dimensional plasmon, which is
produced by electrons moving along the thin metallic cylinder (wire) with both
longitudinal and transverse electric fields outside the cylinder. It has
a complicated dispersion law and its frequencies cover all the range from 
zero to the effective plasma frequency $\omega_p \sqrt{f/\epsilon_m}$. 
Such a plasma mode
 can be excited by s-polarized EMW and it contributes to the longitudinal
component of the dielectric tensor $\epsilon_{\|}$, which is a function of the 
wave vector ${\bf k}$, because the plasma frequency depends on  ${\bf k}$.
Another important plasma mode is a surface mode with the frequency
 $\omega_p/\sqrt {\epsilon_m+1}$. This mode is almost independent of ${\bf k}$.
It is excited by p-polarized EMW and it contributes to the perpendicular
component of the dielectric tensor $\epsilon_{\perp}$, which is 
$k$-independent.

These two plasma modes generate five different branches of the EMW, which cover
all the frequency range except very narrow omnidirectional gap in the vicinity
of the frequency of the surface plasmon. The modes can be classified as two
ordinary waves ($E_z=0$) and three extraordinary waves ($E_z\neq 0$). This
classification comes from the uniaxial dielectric crystal, but the amount of
modes in the metallic mesh is larger because the Fresnel equation becomes
non-linear due to the frequency and the spatial dispersions of the
dielectric tensor. 

The theory of reflection and refraction of the EMW in such a system becomes a
special problem. We have described a 
general approach to this problem and solved
it for the case of particular plane of incidence. 
The reflection coefficient has
a maximum in the narrow frequency interval  of the omnidirectional gap.

Finally, this system is a good example of the ``transparent metal'', 
which has a
metallic conductivity in z-direction and is transparent almost in all frequency
range where condition $\omega \tau \gg 1$ is fulfilled.

\section{Acknowledgment}

We are grateful to V. Vardeny and A. Zakhidov for introducing
us to a marvelous physics of photonic crystals and 
to V. M. Agranovich for very useful discussion of the light
refraction in a system with the spatial dispersion of
dielectric constant.

The work has been supported by NIRT Program of the 
NSF under Grant No. DMR-0102964.

\newpage

\end{document}